\documentclass{emulateapj}

\slugcomment{arxiv preprint---not yet submitted}

\shorttitle{Tuna genome sequences}
\shortauthors{McWilliam et al.}

\begin{document}

\title{A draft genome assembly of southern bluefin tuna {\em Thunnus maccoyii}}

\author{S. McWilliam}
\affil{CSIRO Agriculture, 306 Carmody Road, St. Lucia 4067, Australia}

\author{P. M. Grewe}
\affil{CSIRO Oceans and Atmosphere, Castray Esplanade, Hobart, Tasmania, Australia}

\and

\author{R. J. Bunch and W. Barendse\altaffilmark{1}}
\affil{CSIRO Agriculture, 306 Carmody Road, St. Lucia 4067, Australia}

\email{bill.barendse@gmail.com}

\altaffiltext{1}{School of Veterinary Science, The University of Queensland, Gatton 4343, Australia.}

\begin{abstract}
\noindent
Tuna are large pelagic fish whose populations are close to panmixia.
In addition, they are threatened species, so it is important
for the maintenance and monitoring of genetic diversity that genetic
information at a genome level be obtained.
Here we report the draft assembly of the southern
bluefin tuna genome and the collection of genome-wide
sequence data for five other tuna species.
We sampled five tuna species of the genus {\em Thunnus}, the
northern and southern bluefin, yellowfin, albacore, and bigeye, as
well as the skipjack ({\em Katsuwonis pelamis}), a tuna-like species.
Genome assembly was facilitated at k-mer$=$25 while k-mer$=$51
generated assembly artefacts.
The estimated size of the southern bluefin tuna genome was 795 Mb.
We assembled two southern bluefin tuna individuals 
independently using both paired end and mate pair sequence.
This resulted in scaffolds with N50$>$174,000 bp and 
maximum scaffold lengths$>$1.4 Mb.
Our estimate of the size of the assembled genome was the scaffolded sequences
in common to both assemblies, which amounted to 721 Mb of the 
795 Mb of the southern bluefin tuna genome sequence.
Using BLAST, there were matches between 13,039 of 14,341 (91\%)
refseq mRNA of the zebrafish {\em Danio rerio} to the tuna
assembly indicating that most of a generic fish transcriptome was
covered by the assembly.

\end{abstract}

\keywords{genome; DNA; tuna}

\section{Introduction}

The annual wild tuna harvest is of the order of 4 million tonnes
per year and, with the decline or stringent management of
fisheries, monitoring of stocks and movement of animals is
important to the long term sustainability of tuna fisheries
\citep{Campbell00,MacKenzie09,Pacific14}.
True tuna are the species in the genus {\em Thunnus}, such as
bluefin (Atlantic ({\em T. thynnus}), northern Pacific ({\em T. orientalis}),
and southern ({\em T. maccoyii})), yellowfin ({\em T. albacares}), albacore
({\em T. alalunga}), and bigeye ({\em T. obesus}).
Many of these species are threatened or critically endangered and
are being replaced as a food source by related `tuna' fish
such as the skipjack ({\em Katsuwonis pelamis}).

Genetic studies of these species have a long history and have
shown that, as expected, the populations of these species are
close to panmixia.
In general, pelagic marine fish have low F$_{ST}$ values across
ocean-wide samples and tuna species are no exception
\citep{Elliott92,Ward95,Takagi99,Ward00,Appleyard01,Terol02,Martinez06,Rooker07,Lowenstein09,Riccioni10,Montes12}.
Such panmixia makes monitoring difficult if there are only a small number of
DNA markers and so it is important that a large number of polymorphisms are
available for these species.

One option for generating large numbers of polymorphisms is to
use short read genome sequencing \citep{Bentley08s}.
Initial genome sequencing of a tuna species has occurred, for the northern
Pacific bluefin tuna and the existence of a preliminary assembly has
been reported \citep{Nakamura13}.
Single nucleotide polymorphisms (SNP) for monitoring populations have been
identified in yellowfin \citep{Grewe15,Pecoraro16} and DNA based methods for
identification of tuna species have recently been published
\citep{Takashima14,Santaclara15}.
Next generation sequencing has also recently been used to
obtain the complete mitochrondrial sequence of two tuna species
\citep{Chen16,Li16a}, to start the identification of
microsatellites across the Scombridae \citep{Nikolic15},
and to develop functional genomics tools for the study of
tuna in aquaculture \citep{Yasuike16}.
Here we report additional genome sequence data for
a set of tuna species particularly
the southern bluefin tuna (SBT).
Our main objective is to provide genome sequence information
that might contribute to the identification of SNPs,
to make tools to monitor tuna fisheries, and to generate
public data to improve tuna genome assemblies.
Other objectives are to quantify, where possible, the genetic
differences between species revealed through genome sequencing and to provide
a resource for the further genomic study of pelagic ocean fish.
In this study we report the collection of tissue samples,
of genome sequence for the six species, and the draft
assembly of the southern bluefin tuna genome.

\section{Materials and methods}

\subsection{DNA samples}

Tissue samples of tuna fish were obtained from wild caught tuna as
part of commercial, authorised fishing quotas.
The various tuna fish are easy to discriminate from each other
using colour, length, and shape of dorsal and pectoral fins,
colour of the back and sides, size of eyes, presence or absence of body
stripes \citep{Collette01,Itano05,Joshi12} and we visually checked
the species identity of each individual
(see www.iotc.org/science/species-identification-cards checked
7 July 2016).
Samples were taken post mortem, chilled, and then frozen once they
reached the laboratory.
DNA for genome sequence was purified using a modified CTAB
(cetyl trimethylammonium bromide) method \citep{Doyle87}.

DNA for genome sequencing was checked for quality control using agarose
gel electrophoresis, followed by quantification, DNA fragmentation,
and size selection using a Nanodrop, a Covaris sonicator,
and Qubit-HS DNA kit, as described \citep{Barendse15}.
DNA quality was determined by running a subsample 
on a 1\% 1 X TAE agarose gel to check for DNA smears due
to DNA degradation and the 260:280 nm absorbance ratio on
the Nanodrop was required to be between 1.8 and 2.0.
Each sample was normalized to 20~ng/$\mu$l and 55~$\mu$l was fragmented
to an average insert length of 300 to 400 bp using a Covaris
S220 ultra-sonicator (Woburn, MA, USA).

\subsection{DNA libraries and short sequence reads}

DNA libraries for sequencing were prepared from at least 1.0
$\mu$g of genomic DNA of each individual using the Illumina
TruSeq DNA Sample Prep Kit v2-Set A kit following the manufacturer’s
instructions (Illumina Australia P/L, Scoresby, Vic).
A TruSeq PE Cluster Kit v3 was used to generate sequencing
clusters using the Illumina cBot on Flow Cell v3.
100 bp paired end (PE) reads were obtained using the Illumina
TruSeq SBS Kit v3 - HS kit following the manufacturer’s instructions
on the SQ module of an Illumina HiScan instrument or a HiSeq instrument.
Individuals were multiplexed at 8 samples per lane.
Two of the southern bluefin samples were
sequenced using the Illumina Nextera Mate Pair Library Kit (MP)
following the manufacturer's instructions at insert sizes of
2~kb, 5~kb, and 8~kb.

\subsection{Sequence analysis and genome assembly}

Initial quality control of DNA sequence reads followed a
standard genome analysis pipeline \citep{Barendse15}, except
that there was no reference genome sequence for any tuna species.
Initial quality control consisted of removing sequences that
failed chastity filtering using Illumina Casava 1.8.
Sequences passing chastity filtering were quality trimmed to
remove bases below phred Q20.
Using the fastq trimmer module of the Fastx toolkit
(http://hannonlab.cshl.edu/fastx\_toolkit/command-line.html)
sequences less than 50~bp following trimming were removed
from further analysis.
The k-mer distribution of the reads was evaluated using Jellyfish
\citep{Marcais11} and used to estimate the genome size of tuna.
All short sequence reads of tuna fish that survived
trimming were assembled using SOAPdenovo2 v2.04 \citep{Li10,Luo12}
at the CSIRO High Performance Cluster at two different k-mer sizes,
25 and 51.
A small k-mer size was examined to cope with the potentially
large number of DNA polymorphisms that could be encountered
in pelagic marine species with large population sizes.
Contigs reported as part of the assembly had lengths greater than 100 bp.
The assembly settings for the PE reads were
average insert size 300 bp, maximum read length 100 bp,
asm flags 3, pair number cutoff 3, and map length 32.
Assembly settings for the MP reads were average insert size 3000 or 5000,
asm flags 2, pair number cutoff 5, and map length 35.
To form scaffolds, contigs greater than 200 bp were used.
Using the draft SBT genome assembly as a reference, we compared the RNA
or contigs of the other species to the SBT using BLAST \citep{Altschul90}.
Fastq format reads
of the Pedrabranca3 (PDB3) and Pedrabranca5 (PDB5) individuals
were lodged in the Short Read Archive associated with BioProject 
ID PRJNA327759 which will continue to be updated as the project
progresses.

\section{Results}

\subsection{Sequence collection and genome size estimate}

In total, there were 15 southern bluefin, 3 northern bluefin,
3 albacore, 3 bigeye, 4 skipjack, and 2 yellowfin individual
tuna samples that passed DNA quality control and were sequenced
using PE methods.
The number of reads from each individual after
merging and trimming is shown in
Table~1, indicating for each species
that there was more than 27X genome coverage.
GC content of the genomes was similar, from 39.8\% to 40.8\%.
Two of the SBT animals, PDB3 and PDB5, were
chosen for additional PE and for MP sequencing based
on the amount and quality of DNA.
We assembled sequence reads of PDB3 by itself, of reads of PDB5 by itself,
of reads of PDB3 and PDB5 combined, and
of reads of all available SBT animals combined.
When assembling reads of PDB3 or PDB5 by itself,
only the MP sequence reads of a single individual was used, but
for SBT assemblies using reads of multiple individuals,
combined MP sequence of PDB3 and PDB5 was used.
The 8 kb MP reads proved to have low complexity
and were excluded from assemblies.

We estimated the genome size of tuna to be 795 Mb based on the k-mer
distribution. A plot of the k-mer distribution is shown in Figure~1.
This shows the count of 17-mers on the horizontal axis where 40 represents
a 17-mer that occurs 40 times.
The vertical axis shows the sum of the count of 17-mers
so that the curve of the
graph shows the total number of 17-mers that occurred a particular
number of times.
For example, the total number of all 17-mers that occurred 40 times
is 13,211,979.
For a k-mer length of 17, there was read coverage of 44 for the genome and a
peak k-mer depth of 37.

\subsection{Assembly information for species lacking MP reads}

Although we did not have MP sequence for five of the species,
we attempted to assemble the reads because the contigs would
be useful for discovering DNA variants. 
In addition, once the SBT was
assembled using MP reads, the contigs of the
other five species obtained in this process
could be aligned to that assembly 
better than merely matching heterologous fastq reads to the SBT assembly.
The information on these five species is shown in Table~2 which
includes part of the SBT dataset as a comparison.
The SBT sequences in this comparison are a mixture of all the PE sequences
of PDB3 and PDB5, giving a similar dataset to the other species.
In these comparisons, the SBT PE only assembly showed no
special characteristics compared to the others
and had similar statistics to the northern bluefin tuna.
Between 67\% and 74\% of a tuna genome was covered by contigs of 
longer than 100 bp (Table~2).
Estimates of coverage, N50, and maximum length of contigs or
scaffolds did not appear to be strongly related to amount or
type of data.
The scaffolds generated in this process showed a 
several fold improvement in N50 and maximum length
by a factor of two to five compared to contigs before
scaffolding occurred.
These multi-contig scaffolds covered less of the genome
than the full list of contigs, and in the case of the skipjack,
the coverage by scaffolds was markedly less than the true tuna species.

\subsection{Draft assembly of the southern bluefin tuna}

We constructed draft SBT assemblies using both PE and
MP sequence using several datasets to determine the best
approach for that species (Table~3).
In particular we wanted to know whether
increasing the number of reads and number of individuals
would lead to superior results for this species.
Although PDB3 did but PDB5 did not quite
reach 30X coverage of the tuna genome, both showed massive reductions in
scaffold number and several order of magnitude increases in 
scaffold length compared to lengths of contigs prior to scaffolding.
This occurred whether the k-mer length was 25 or 51.
As more reads from different individuals were added
to the assembly, contig number
increased, N50 and maximum lengths decreased, and scaffolds
became shorter.
This feature became worse at longer k-mer lengths.
The likely cause is increased genetic variability in the dataset
as more individuals were added.

Longer k-mers did not improve the sequence and may have
decreased the quality of the assemblies.
Although the number of contigs and the apparent coverage of the tuna genome
is greater with k$=$51 (Table~3), this coverage exceeds the known length of the
tuna genome by 18\% or more than 140 Mb. 
Two factors may contribute to this result.
Firstly, with shorter k-mers and a threshold of
100 bp for inclusion in the list of counted contigs, more short contigs
are excluded with k$=$25.
These short contigs have reduced sequence variability
compared to longer contigs and may represent segments of the
same repeat structure (Figure~2).
Contigs such as those containing mononucleotide runs are excluded
at shorter k-mer lengths.
As contig length increases, the major change in the graph in Figure~2
is that the minimum diversity values increase,
and at short contig lengths the low diversity values
are driven by poly(A), poly(C), poly(G), and poly(T) segments.
Secondly, the distribution of contig length by contig frequency
showed a clear artefact in the distribution of the
k$=$51 contigs for both PDB3 and PDB5 (Figure~3).
This artefact would inflate the coverage by contigs of size 100 to 200 bp.
Further analyses of the draft assemblies were restricted to k$=$25.

To estimate the coverage of the SBT tuna genome, first we counted the
length of each assembly (Table~3).
This showed values for scaffolds that exceeded the length of the SBT genome.
A BLAST analysis of the PDB3 scaffolds against the PDB5
scaffolds showed 725 Mb in common, i.e., 
the intersection of the two draft assemblies was 725 Mb.
This is likely to be an overestimate of the 
coverage of the genome because there would be some
short contigs that only match for 33 bp and at that level of similarity
they may be referencing different parts of the genome.
Excluding BLAST matches of $\leq$ 50 bp resulted in joint coverage of 721 Mb
of the 795 Mb SBT genome.
These BLAST analyses showed that alignment of the two assemblies
filled sequence missing from one or the other assembly (Figure~4),
sequence represented as runs of Ns in one or the other assembly.
There were 143,134 segments where the BLAST alignment filled
in missing sequence in this way.
An additional 1,402 matching segments had gaps in the
same locations in both assemblies.

\subsection{Gene content of the draft SBT assembly}

A BLAST analysis was used to quantify the RNA coverage of the assembly
using the RNA complement of the zebrafish {\em Danio rerio}.
The RNA sequences contained in GCF\_000002035.5\_GRCz10\_rna.fna
were matched to either assembly using BLAST.
Most scaffolds showed a single match but some scaffolds in both assemblies
showed more than 50 matches.
The distribution of BLAST matches is the same for PDB3 compared to
PDB5 (Figure~5).
As increased stringency is used, fewer RNA sequences matched and
the overlap between PDB3 and PDB5 decreased,
as expected.
Nevertheless, at a BLAST probability threshold of 1-e15 (Table~4),
45,135 of the RNAs were located to either of the two assemblies,
of which 30,868 were found on both assemblies.
At the more stringent threshold of 1-e40
28,461 of the RNA were located to either assembly of which 11,160
were found on both assemblies.
The total number of RNA items tested from {\em D. rerio} was 54,437.
For reference sequences beginning with `NM\_', taken
to be the refseq mRNA complement, there were a
total of 14,341 in the {\em D. rerio} set of which 13,039 were located
to either of the assemblies at a threshold of 1-e15 and
8,222 were located to either of the
assemblies at a theshold of 1-e40 (Table~4).
In conclusion, a very large part of the transcriptome of a generic
fish genome was located either totally or in part to the SBT genome.

An examination of one of the large scaffolds, scaffold269 of the PDB3
assembly, showed greater synteny between tuna and fugu ({\em Takifugu rubripes})
than tuna and zebrafish in that scaffold.
Scaffold269 was chosen because it matched to 18 RNAs.
Genome locations were sought in the zebrafish genome for these genes.
The locations were Dre15: sgcg, trappc4, sacs;
Dre11: foxg1b, foxg1c;
Dre10: rps25, XM\_005172584.2, XM\_009305610.1, XM\_699844.6;
Dre18: cfap45, pglyrp5, snrpd2, polr2i, c5ar1, vaspb, XM\_001336435.5, XM\_001340960.3; and
Dre13: foxg1d.
Not all of the {\em D. rerio} RNA were found in fugu using a search on names.
Of those that were, the locations were Tru11: sgcg, trappc4, sacs, foxg1b, rps25, cfap45, snrpd2, polr2i, foxg1d;
Tru4: foxg1c;
not found in Tru: pglyrp5, c5ar1, vaspb, XM\_005172584.2, XM\_009305610.1, XM\_699844.6, XM\_001336435.5, XM\_001340960.3.

\section{Discussion}

We report the genome sequencing of six tuna species
and the draft genome assembly of the southern bluefin tuna.
Our estimate of the tuna genome was 795 Mb, which is consistent 
with the 800 Mb genome size estimated using
other methods \citep{Hardie04,Gregory07}.
Although it is often recommended that assemblies be performed
with larger kmers, we found that at k$=$51 artefacts were generated
in smaller contigs which inflated statistics of genome coverage.
Although five of the six genomes were analysed primarily to
obtain DNA variants for polymorphism discovery, MP sequence
was obtained to generate a {\em de novo} assembly of the 
southern bluefin tuna.
The {\em de novo} assemblies of the SBT
resulted in scaffolds with N50$>$174,000 bp
and maximum scaffold lengths$>$1.4~Mb.
Comparison of the PDB3 and PDB5 assemblies showed that 721 Mb of the 795 Mb
SBT genome was aligned to scaffolds and could find a matching sequence in 
the other alignment.
Examination of the matching scaffolds showed 143,134 segments
where PDB3 sequence filled in a poly(N) sequence of PDB5
and {\em vice versa}.
Of the refseq mRNA of {\em Danio rerio}, 13,039 of 14,341 sequences
were located to either PDB3 or PDB5, with 10,464 sequences located
to both PDB3 and PDB5.
This confirmed that a large proportion of the coding sequence of the SBT
was contained in the scaffolds of the SBT assembly.
Further annotation of the gene complement of the SBT genome
will require a bioinformatics community effort.
Preliminary examination of synteny in a scaffold with 18
different coding sequences showed that synteny of those genes
was highly conserved to fugu but there was
less conservation of synteny observed to the zebrafish.
Further analyses of synteny will require detailed mapping of
elements and confirmation of homology, which is
beyond the scope of this communication.
Such analyses will be required to demonstrate whether the
conservation of synteny 
continued to be greater for the pair of tuna and fugu compared to the 
pair of tuna and zebrafish.
Further analyses of the DNA variants uncovered in this study
will be performed and reported in a subsequent manuscript.

We found that with increasing stringency that an RNA sequence would
match to one rather than both assemblies.
This is unlikely
to be due to genetic variability but rather due to places
in the assembly where one assembly has poly(N) where the other
assembly has reportable sequence. 
Since the sequence that is sampled is partly dependent upon
chance, the presence of reportable sequence in one animal
but poly(N) in another indicates that even with around 30X
genome coverage there are many areas of the genome that 
are counted as covered but which have short sections of
unknown bases.
Our results suggest that it is preferable to keep the
genome data for each animal separate but that it may be 
advantageous to fully sequence more than one individual
and compare the assemblies to fill in missing sequence.
This may be because of the random nature of sequence capture,
genome segments that may be well captured in one animal during one
sequence run may not be well captured at some other time
in a different animal.
Our data do not say whether it is preferable to sequence
one individual to 60X coverage compared to two individuals each
to 30X coverage with separate assemblies for the two
individuals. 
However, where the description of DNA variability is important
then the sequencing of two individuals each to 30X coverage
will yield more DNA variants than sequencing one individual
to 60X coverage.
Our results suggest that future analyses of the other
five tuna species should try to assemble each individual
separately, compare the resulting assemblies using BLAST
or some other tool, and then compare the resulting consensus
assembly to the SBT draft assembly to construct larger
scale scaffolds of those tuna genomes.
This method should be used even if the total number of
reads per individual does not reach the 30X level of sequence
coverage.

To quantify the sequence diversity of short contigs
we used the Shannon Information index because it
is a standard method.
The plot in Figure 3 clearly shows that as contig
length increases so the occurrence of contigs with
low information declines, low information in this case
meaning mononucleotide runs of DNA bases or similar low diversity
patterns.
There are contigs where the maximum index value can be found
for both short and longer contigs, but 
it is only among the short contigs that very low index
values are found.
These correspond to contigs that have for example 26 out of
26 sequential adenine nucleotides.
Consequently, as contig length increased the
mean index value tracked upwards
towards the maximum value and the standard deviation
reduced as the extremely small index values disappeared.
The data show that much of this low diversity
occurs for contigs below 50 bp in size, and so setting
k-mer$=$25 acts as a way of flushing out low diversity
contigs.

The BLAST alignment of the PDB3 and PDB5 assemblies
totals 721 Mb and we take that as our estimate of
genome coverage of the SBT genome.
This estimate is very conservative but we see no point
in including small contigs that are not strongly placed
or which might represent repeat sequences that are
located in multiple regions of the tuna genome.
The accurate localization of such repeats, whether they are
telomeres, centromeres, short interspersed nuclear
elements, or other 
low complexity regions, requires special
techniques beyond the scope of this study.
To include them in the count of the percent of
the genome covered would give a false impression.
Much further research 
is needed to complete the genome 
assembly of a tuna species and to make it as useful as possible.
The following still needs to be done: 1) a survey of the
DNA variants and their frequency distribution; 
2) a cataloging of the coding sequences and the many decisions
on homology that need to be made;
3) the identification of
the full transcriptome of tuna; 4) the assignment
of scaffolds to chromosomes; 5) the identification of
promotor, enhancer, and other elements relevant to 
gene expression; 6) and the identification of the
genome sections association with sex determination.
While those are a wish-list for the future,
the draft assembly presented here is a
step forwards towards that goal.

\acknowledgments

{\em Acknowledgements}---We thank CSIRO for support and
access to the High Performance Computing Cluster and to CSIRO
and the University of Queensland for access to their library services.
We thank Ross Tellam and Brian Dalrymple who read the manuscript and made
comments which improved the text.
Both noted the absence of Biology.
We thank Blair Harrison and Janette Edson for technical assistance.
Fish samples were obtained or coodinated by PMG.
DNA was extracted by PMG and RJB.
Sequencing libraries were prepared by RJB.
Quality control analysis of fastq files for sequence analysis 
was performed by SM and RJB.
Assembly of genomes and analysis of data was performed by WB and SM.
Manuscript drafted by WB with comments from the other authors.
The manuscript passed CSIRO internal review 6 July 2016.

\bibliographystyle{authordate1}

\clearpage

\begin{figure}
\plotone{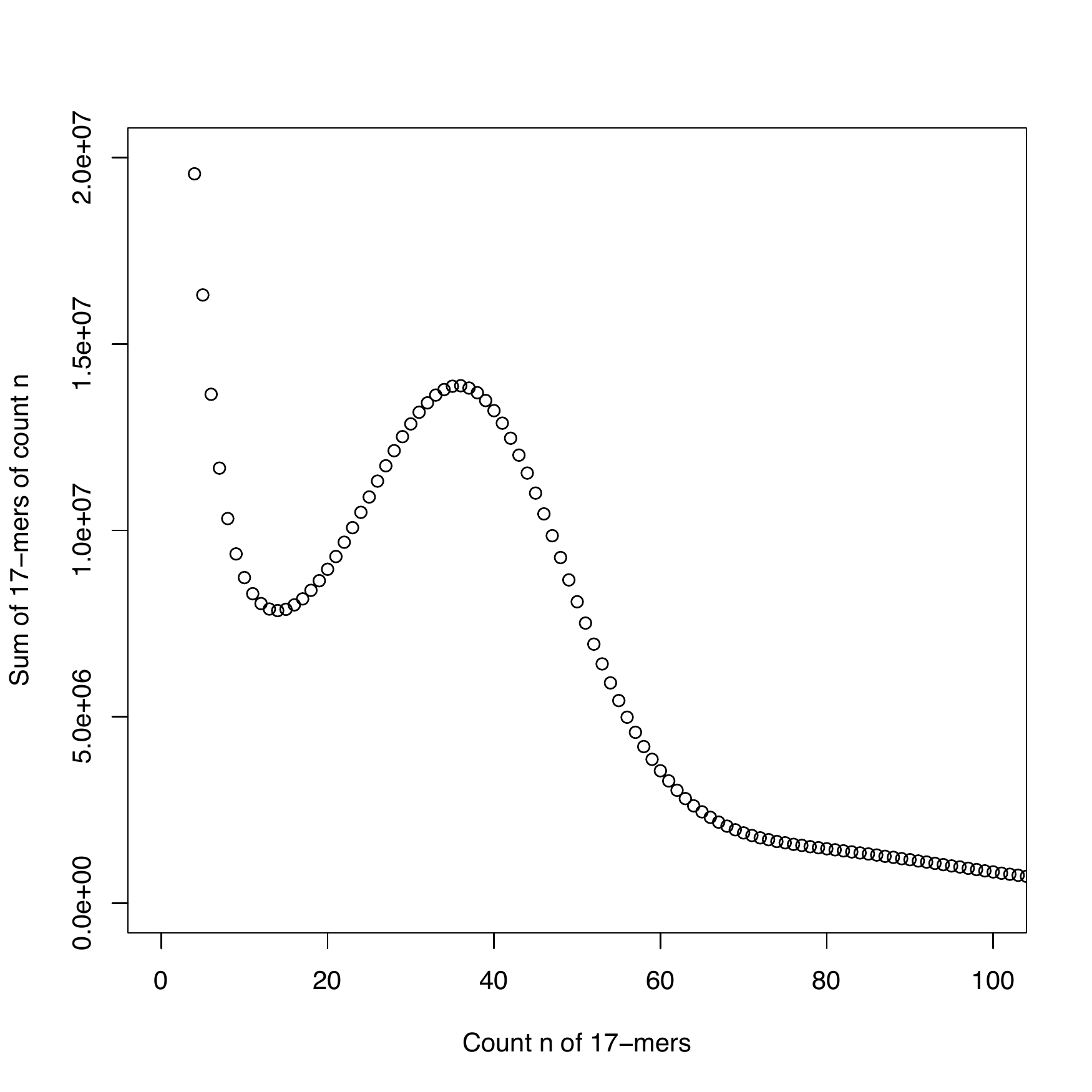}
\caption{The k-mer distribution of 17-mers of southern bluefin tuna yielding a genome size estimate of 795 Mb}
\end{figure}

\clearpage

\begin{figure}
\plotone{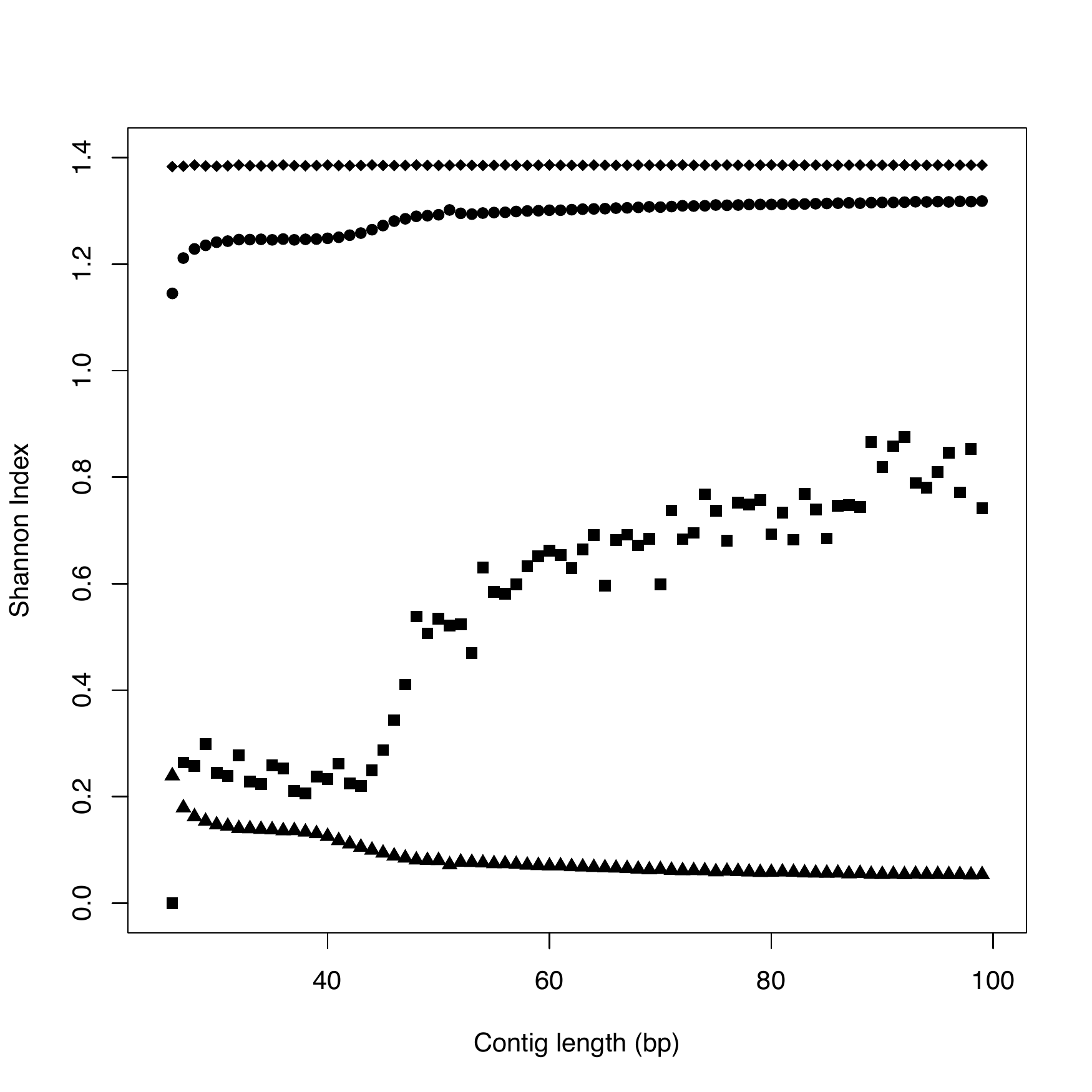}
\caption{The increase in contig sequence diversity with increased contig length. Diversity is expressed as the Shannon Information Index. Minimum diversity is shown with squares, maximum diversity with diamonds, the mean with circles, and the standard deviation with triangles}
\end{figure}

\clearpage
\begin{figure}
\plotone{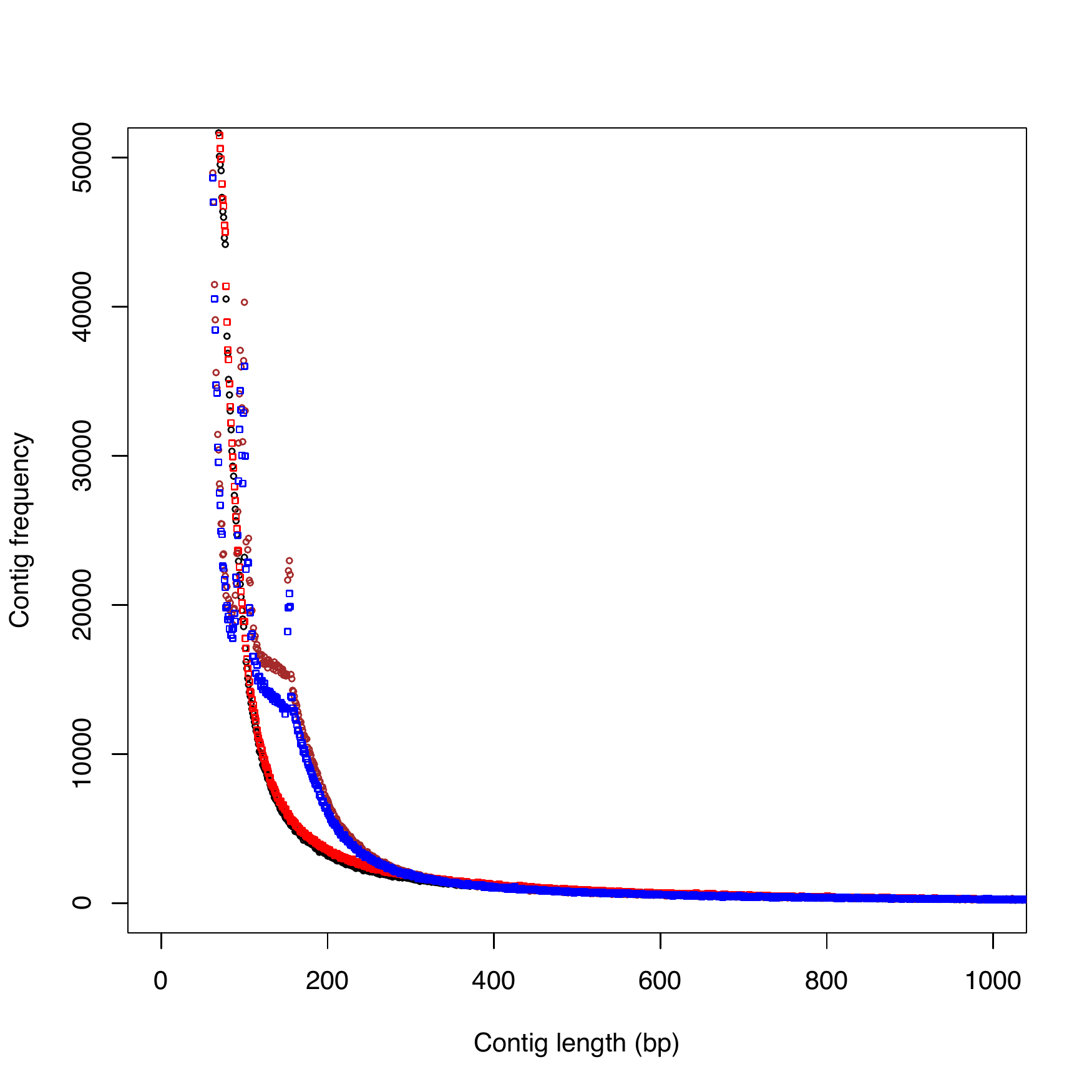}
\caption{The size distribution of contigs in base pairs for PDB3 k$=$25 in black k$=$51 in brown and PDB5 k$=$25 in red k$=$51 in blue}
\end{figure}

\clearpage

\begin{figure}
\plotone{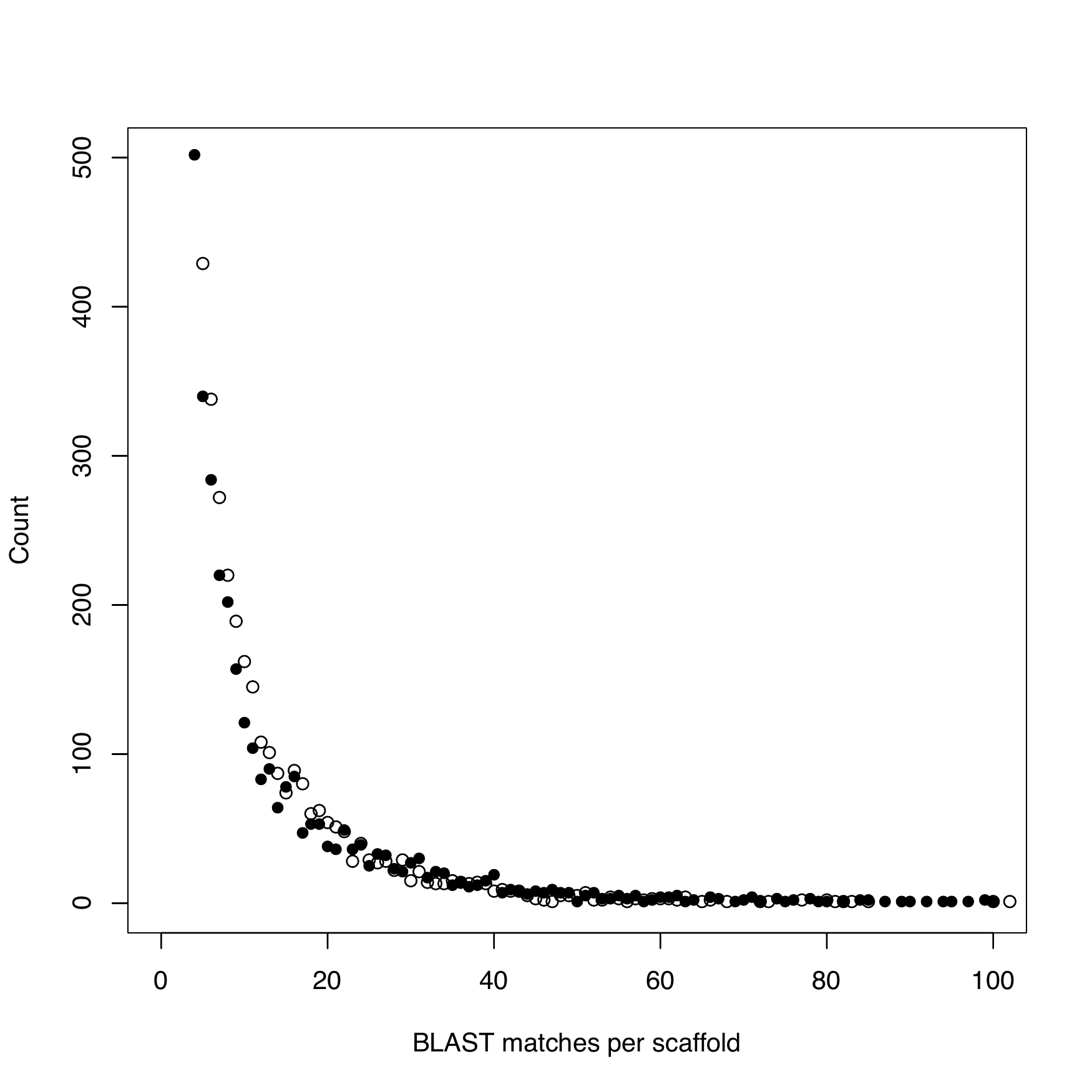}
\caption{The distribution of number of matches per scaffold for PDB3 (open circles) and for PDB5 (closed circles) for {\em Danio rerio} RNA}
\end{figure}

\clearpage

\begin{figure}
\plotone{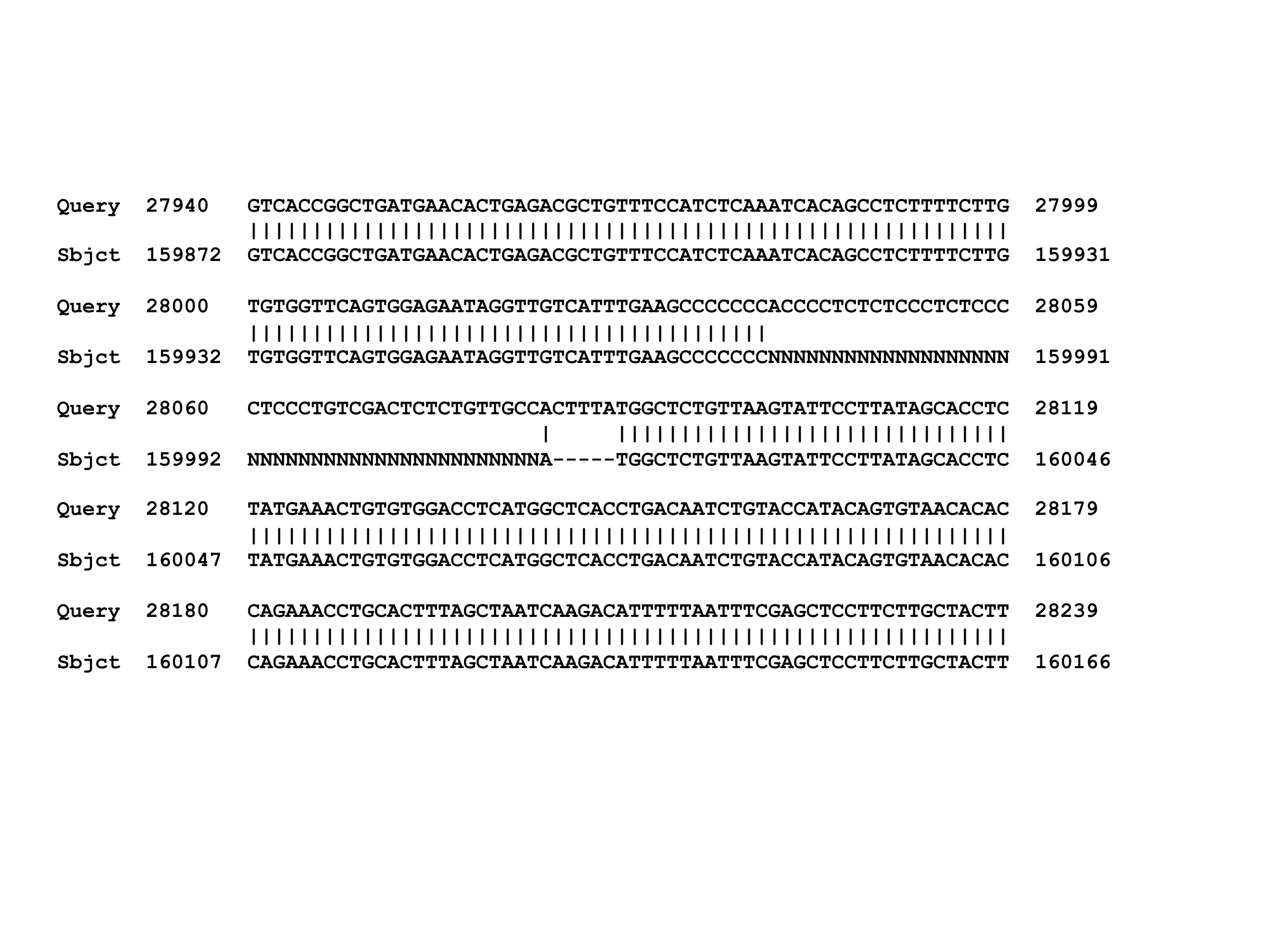}
\caption{BLAST alignment between part of a PDB3 and PDB5 scaffold where sequence complementarity increases coverage}
\end{figure}

\clearpage

\begin{deluxetable}{lrrrrc}
\tabletypesize{\scriptsize}
\tablecaption{Genome sequence used for tuna genome assembly. \label{tbl1}}
\tablewidth{0pt}
\tablehead{
\colhead{Species} & \colhead{N} & \colhead{GC percent} & \colhead{PE reads} & \colhead{MP reads} & \colhead{PE Coverage} 
}
\startdata
albacore (ALB) & 3 & 40.26 & 533,707,152 & - & 66X \\ 
bigeye (BET) & 3 & 39.84 & 220,316,910 & - & 28X \\ 
northern bluefin (NBT) & 3 & 39.99 & 352,965,874 & - & 44X \\ 
skipjack (SJK) & 4 & 40.82 & 734,059,014 & - & 91X \\ 
southern bluefin (SBT)\tablenotemark{a} & 13 & 40.72 & 1,694,446,949 & - & 213X \\ 
SBT PDB3\&PDB5 & 2 & 39.94 & 479,524,357 & 1,266,757,744 & 60X \\ 
yellowfin (YFT) & 2 & 39.81 & 240,409,742 & - & 30X\\ 
\enddata
\tablecomments{The number of reads reported is for merged, trimmed, and paired
reads, representing the reads used in the assembly, which are a subset of the
reads collected for each species}
\tablenotetext{a}{Total SBT is 2,173,971,306 reads, which is the sum of these
reads and the SBT Pedrabranca reads}
\end{deluxetable}

\clearpage

\begin{deluxetable}{lcccccccc}
\tabletypesize{\scriptsize}
\tablecaption{Tuna Assembly Statistics. \label{tbl2}}
\tablewidth{0pt}
\tablehead{
 & Contig & & & & Scaffold\tablenotemark{a} & & & \\
Species & Number & Coverage & N50 & Max & Number & Coverage & N50 & Max \\
\vspace{-0.2cm}  & & (Gb) & (bp) & (bp) & & (Gb) & (bp) & (bp) \\
}
\startdata
albacore & 1,826,784 & 545 & 398 & 7,187 & 176,177 & 343 & 1,406 & 43,393 \\
bigeye & 1,533,456 & 606 & 640 & 15,453 & 147,654 & 454 & 3,207 & 65,966 \\
northern bluefin & 1,741,562 & 575 & 479 & 11,837 & 174,297 & 391 & 1,837 & 45,910 \\
skipjack & 1,812,717 & 422 & 271 & 5,119 & 134,886 & 194 & 608 & 45,518 \\
southern bluefin\tablenotemark{b} & 1,756,506 & 590 & 489 & 8,422 & 174,892 &
407 & 1,973 & 44,602 \\
yellowfin & 1,584,634 & 607 & 616 & 15,538 & 157,596 & 446 & 2,760 & 63,041 \\
\enddata
\tablecomments{These assemblies were constructed without MP sequences and with k-mer$=$25}
\tablenotetext{a}{Number of scaffolds consisting of more than one contig}
\tablenotetext{b}{SBT PDB3 and PDB5 PE sequences only, k$=$25}
\end{deluxetable}

\clearpage

\begin{deluxetable}{lrrrrrrrr}
\tabletypesize{\scriptsize}
\tablecaption{Southern bluefin tuna alternative assembly statistics. \label{tbl3}}
\tablewidth{0pt}
\tablehead{
 & Contig & & & & Scaffold\tablenotemark{a}& & & \\ 
Assembly & Number & Coverage & N50 & Max & Number & Coverage & N50 & Max \\
\vspace{-0.2cm} & & (Gb) & (bp) & (bp) & & (Gb) & (bp) & (bp) \\
}
\startdata
PDB3-pe\tablenotemark{b} & 1,523,729 & 629 & 720 & 16,076 & 162,973 & 459 & 2,708 & 59,033 \\ 
PDB3-mp\tablenotemark{c} & 1,523,729 & 629 & 720 & 16,076 & 12,828 & 890 & 174,777 & 1,412,496 \\ 
PDB3x5-mp\tablenotemark{d} & 1,523,729 & 629 & 720 & 16,076 & 12,827 & 945 & 222,019 & 1,707,042 \\ 
PDB3-mp51\tablenotemark{e} & 2,630,814 & 938 & 709 & 37,065 & 14,814 & 930 & 147,567 & 1,412,291 \\ 
PDB5-pe\tablenotemark{f} & 1,625,821 & 627 & 626 & 13,874 & 169,576 & 455 & 2,538 & 50,265 \\ 
PDB5-mp\tablenotemark{g} & 1,625,821 & 627 & 626 & 13,874 & 11,740 & 963 & 245,122 & 2,315,601 \\ 
PDB5x3-mp\tablenotemark{h} & 1,625,821 & 627 & 626 & 13,874 & 13,215 & 879 & 181,542 & 1,323,743 \\ 
PDB5-mp51\tablenotemark{i} & 2,472,157 & 920 & 768 & 30,018 & 11,139 & 998 & 243,461 & 2,056,612 \\ 
PDB3\&PDB5-pe\tablenotemark{j} & 1,756,506 & 590 & 489 & 8,422 & 174,892 & 407 & 1,973 & 44,602 \\ 
PDB3\&PDB5-mp\tablenotemark{k} & 1,756,506 & 590 & 489 & 8,422 & 244,152 & 298 & 922 & 9,801 \\
PDB3\&PDB5-mp51\tablenotemark{l} & 4,629,720 & 1,089 & 280 & 20,743 & 223,853 & 340 & 754 & 22,277 \\
All-pe\tablenotemark{m} & 1,813,209 & 439 & 285 & 4,957 & 144,536 & 228 & 801 & 27,046 \\
All-mp\tablenotemark{n} & 1,813,209 & 439 & 285 & 4,957 & 193,004 & 178 & 522 & 6,402 \\
All-mp51\tablenotemark{o} & 10,794,475 & 1,516 & 112 & 14,818 & 18,716 & 21 & 86 & 14,767 \\
\enddata

\tablecomments{There are significant differences in the quality of these
assemblies by changing the k-mer size and the data set}
\tablenotetext{a}{Number of scaffolds consisting of more than one contig}
\tablenotetext{b}{PDB3 PE sequences only, k$=$25}
\tablenotetext{c}{PDB3 including MP sequences, k$=$25}
\tablenotetext{d}{PDB3 using PDB5 MP sequences, k$=$25}
\tablenotetext{e}{PDB3 including MP sequences, k$=$51}
\tablenotetext{f}{PDB5 PE sequences only, k$=$25}
\tablenotetext{g}{PDB5 including MP sequences, k$=$25}
\tablenotetext{h}{PDB5 using PDB3 MP sequences, k$=$25}
\tablenotetext{i}{PDB5 including MP sequences, k$=$51}
\tablenotetext{j}{PDB3 and PDB5 PE sequences only}
\tablenotetext{k}{PDB3 and PDB5 including MP sequences}
\tablenotetext{l}{PDB3 and PDB5 including MP sequences, k$=$51}
\tablenotetext{m}{All available SBT PE sequences, k$=$25}
\tablenotetext{n}{All available SBT PE and MP sequences, k$=$25}
\tablenotetext{o}{All available SBT PE and MP sequences, k$=$51}

\end{deluxetable}

\clearpage

\begin{deluxetable}{lrrrrrrrrrr}
\tabletypesize{\scriptsize}
\tablecaption{Coverage of Danio rerio RNA on southern bluefin tuna assemblies. \label{tbl4}}
\tablewidth{0pt}
\tablehead{
Assembly & NM\tablenotemark{a} 1-e05 & NM 1-e10 & NM 1-e15 & NM 1-e20 & NM 1-e25 & NM 1-e30 & NM 1-e35 & NM 1-e40 & NM 1-e45 & NM 1-e50
}
\startdata
PDB3 & 44801 & 41539 & 38158 & 34657 & 30903 & 27127 & 23464 & 20027 & 16742 & 13916 \\
PDB5 & 44665 & 41289 & 37845 & 34291 & 30742 & 26954 & 23069 & 19594 & 16249 & 13426 \\
PDB3 and PDB5\tablenotemark{b} & - & - & 10464 & - & - & - & - & 3598 & - & - \\
PDB3 or PDB5\tablenotemark{c} & - & - & 13039 & - & - & - & - & 8222 & - & - \\
PDB3 and PDB5\tablenotemark{d} & - & - & 30868 & - & - & - & - & 11160 & - & - \\
PDB3 or PDB5\tablenotemark{e} & - & - & 45135 & - & - & - & - & 28461 & - & - \\
\enddata
\tablenotetext{a}{number of RNA matching to assembly at BLAST probability 1-e05}
\tablenotetext{b}{The D. rerio mRNA matching to both assemblies at that BLAST probability}
\tablenotetext{c}{The total D. rerio mRNA matching to either assembly at that BLAST probability}
\tablenotetext{d}{The D. rerio RNA matching to both assemblies at that BLAST probability}
\tablenotetext{e}{The total D. rerio RNA matching to either assembly at that BLAST probability}
\end{deluxetable}

\end{document}